\documentclass[11pt,a4paper,DIV11]{scrartcl}
\usepackage{jheppub}
\usepackage[utf8]{inputenc}
\usepackage[T1]{fontenc}
\usepackage{lmodern}
\usepackage[british]{babel}
\usepackage{amsmath}
\usepackage{amssymb}
\usepackage{amsfonts}
\usepackage{color}
\usepackage{float}
\usepackage{stmaryrd}
\usepackage{url}
\usepackage{epsfig}
\usepackage{epstopdf}
\usepackage{graphicx,psfrag}
\usepackage{subfigure}
\usepackage{placeins,bbm}

\newcommand{\dia}{\mathrm{Dirac}}

\title{ Nonperturbative results for two-index conformal windows}
\author[a]{Georg Bergner}
\author[b]{Thomas A. Ryttov}
\author[b]{Francesco Sannino}
\affiliation[a]{Universit\"at Bern, Institut f\"ur Theoretische Physik, Sidlerstr.~5, CH-3012 Bern, Switzerland}
\affiliation[b]{CP$^3$-Origins and the Danish IAS, University of Southern Denmark, 5230 Odense M, Denmark}
\emailAdd{g.bergner@uni-muenster.de,ryttov@cp3.dias.sdu.dk,sannino@cp3.dias.sdu.dk}
\abstract{
Via large and small $N_c$ relations we derive nonperturbative results about the conformal window of two-index theories. Using Schwinger-Dyson methods as well as four-loops results we estimate subleading  corrections and show that naive large number of colors extrapolations are unreliable when $N_c$  is less than about six.  Nevertheless useful nonperturbative inequalities for the size of the conformal windows, for any number of colors, can be derived. By further observing that the adjoint conformal window is  independent of the number of colors we argue, among other things, that: The large $N_c$ two-index conformal window is twice the conformal window of the adjoint representation (which can be determined at small $N_c$) expressed in terms of Dirac fermions;  Lattice results for adjoint matter can be used to provide independent information on the conformal dynamics  of two-index theories such as SU($N_c$) with two and four symmetric Dirac flavors. 
\\[1.5ex]
{Preprint: CP3-Origins-2015-039 DNRF90, DIAS-2015-039}}

\begin{document}
\maketitle

\section{Introduction}

Gauge-Yukawa theories constitute the backbone of our current understanding of fundamental interactions. Yet, despite the deceivingly simple description in terms of the elementary fields and bare interactions very little is known about their dynamics and spectrum. This severely limits the number of theories that can be used to extend, or modify, the standard model of particle interactions. On general grounds it is therefore essential to determine the phase diagram of gauge theories of fundamental interactions\footnote{For example, the recent discovery of  nonsupersymmetric complete asymptotically safe Gauge-Yukawa theories \cite{Litim:2014uca} where elementary scalars are actually needed to make the theory fundamental (i.e. well defined at arbitrary short distances) has widened the horizon of potentially interesting extensions of the standard model and unveiled novel thermodynamical properties \cite{Rischke:2015mea}. Nonperturbative results about the (non) asymptotic safe behaviour of supersymmetric theories appeared in \cite{Intriligator:2015xxa} with earlier studies  reported in \cite{Martin:2000cr}.}.

However, even in the simpler case of gauge theories with only fermionic matter the uncovering of the associated phase diagram presents formidable challenges. Recently analytical \cite{Sannino:2004qp,Dietrich:2006cm} and numerical efforts \cite{Catterall:2007yx,Hietanen:2009zz,DelDebbio:2010hx,DeGrand:2011qd,Appelquist:2011dp,DeGrand:2010na,Fodor:2015zna,Hasenfratz:2015ssa,Athenodorou:2014eua} have been dedicated to determine whether certain gauge theories featuring fermionic matter display large distance conformality. The resulting conformal window plays a central role for a large number of possible extensions of the standard model of particle interactions ranging from near conformal composite dynamics Higgs theories to new dark matter models and even unparticles.  
 
 Using large and small $N_c$ arguments we will show that it is possible to relate the conformal window and dynamics of different gauge theories allowing to fast-forward the investigation of the phase diagram both analytically and via first principle lattice simulations. Specifically we will exploit  large $N_c$ 
 relations  among two-index and adjoint representations \cite{Corrigan:1979xf,Kiritsis:1989ge,Armoni:2004ub} to derive nonperturbative results about the conformal window of these theories. We will also test these relations for the conformal window by using Schwinger-Dyson (SD) methods summarised in \cite{Sannino:2009za} as well as the maximal known order in perturbation theory \cite{Mojaza:2010cm,Pica:2010mt,Pica:2010xq,Ryttov:2010iz}. Thanks to these approximations we will show that naive large $N_c$ extrapolations to low number of colors are unreliable  because one needs to consider at least ${\cal O}(N_c^{-4})$ corrections for describing the conformal window of two-index theories with $N_c$ less than about six.  At order ${\cal O}(N_c^{-1})$ the extrapolation is only reliable down to about $N_c\sim 20$. Nevertheless  $N_c$ independent nonperturbative inequalities can be deduced to significantly constrain the size of two-index conformal windows. By further noting the $N_c$ independence of the adjoint conformal window we will argue that: The large $N_c$ conformal window of two-index theories is twice the conformal window of the adjoint representation in terms of Dirac fermions;  Lattice results, obtained at small $N_c$, for adjoint matter can be used to cross check and predict the conformal dynamics of two-index theories at large and small number of colors. 
 
 The outline of this paper is as follows. In Section \ref{section2} we summarise and further elucidate the large and small $N_c$ relations between theories with adjoint and two-index matter. We then use SD and the maximum known order in perturbation theory to further validate the relations and provide novel insight such as the actual range of applicability of naive large $N_c$ results. Here we will also notice that the size of the adjoint conformal window is effectively $N_c$ independent.  There is only a very mild $N_c$ dependence appearing at four loops which effectively does not change any of our conclusions. We introduce and discuss the implications of nonperturbative inequalities on the size of the two-index symmetric conformal windows in Section \ref{section3}. Here we further show how to use lattice results for the adjoint representation   to cross-check and predict the conformal dynamics of two-index theories such as SU(3) with two Dirac flavors in the sextet representation and offer our conclusions.

\section{Large and small $N_c$ relations and corrections}
\label{section2}
The adjoint theory with $N_{wf}[G]$ Weyl
fermions is planar equivalent to the theory with $N_f[S_2/A_2]=N_{wf}[G]$ Dirac fermions in the two-index
symmetric or antisymmetric representation. Here $G$, $S_2$ and $A_2$ denote the adjoint, two-index symmetric and two-index antisymmetric representations respectively. This fact can easily be seen from a simple counting of the number of fermionic degrees of freedom. For the adjoint representation and for each $N_{wf}[G]$ there are $N_c^2-1$ degrees of freedom. For the two-index symmetric and antisymmetric representations and for each $N_f[S_2]$ or $N_f[A_2]$ there are $N_c(N_c+1)$  or $N_c(N_c-1)$ number of degrees of freedom respectively where we have made sure to count one Dirac fermion as two Weyl fermions and therefore multiplied by a factor of two. Hence in the large $N_c$ limit the number of degrees of freedom equal $N_c^2$ in all three cases and the theory with $N_{wf}[G]$ adjoint Weyl fermions is equivalent to the theory with $N_f[S_2/A_2]$ Dirac fermions in either the two-index symmetric or two-index antisymmetric representation. 

This implies that if we know the conformal window at large $N_c$ for the adjoint theory all we need to do is to re-interpret it as the large $N_c$ conformal window for the two-index symmetric or antisymmetric theories albeit with the number of adjoint Weyl fermions replaced by an identical number of two-index symmetric or antisymmetric Dirac flavors. This  nonperturbative result has direct application to the case of either one or two Dirac flavors, i.e. two or four Weyl fermions, in the adjoint representation, among others. If the former (latter) theory is conformal at large $N_c$ then so is the theory with two (four) Dirac flavors in the two-index symmetric or antisymmetric representation.

Once we depart from the large $N_c$ limit corrections will begin to appear and we will estimate them shortly. For the two-index symmetric and antisymmetric representations these corrections come with opposite sign. The two-index antisymmetric representation corresponds effectively
at finite $N_c$ to a theory with a fewer number of fermionic degrees of freedom while the two-index symmetric representation corresponds effectively to a theory with a larger number of fermionic degrees of freedom. Hence at smaller $N_c$ the boundary of the  conformal window moves to smaller $N_f$ for the two-index symmetric representation and to larger $N_f$ for the two-index antisymmetric representation. We expect the boundary of the conformal window for the adjoint representation to receive only mild corrections at finite $N_c$.

We illustrate this behaviour using first the ladder results \cite{Sannino:2004qp,Dietrich:2006cm} and then via the maximal known order in perturbation theory \cite{Pica:2010xq,Ryttov:2010iz}. The latter is a precise result when the fixed point occurs near the loss of asymptotic freedom because here the fixed point value of the coupling is small.

The ladder approach to determine the lower boundary of the conformal window is summarised in \cite{Sannino:2009za}. In its simplest incarnation it is a truncation of the Schwinger-Dyson (SD) induced gap equation by means of the two loop beta-function.  It was used first in \cite{Sannino:2004qp} for $\mathrm{SU}(N_c)$ gauge theories with two-index matter representation and generalised for arbitrary $\mathrm{SU}(N_c)$ representations in  \cite{Dietrich:2006cm}. Orthogonal and symplectic gauge groups were investigated in \cite{Sannino:2009aw} while exceptional and spinorial representations were studied in \cite{Mojaza:2012zd}.   
We follow the notation in \cite{Dietrich:2006cm} and indicate the upper boundary of the conformal window where asymptotic freedom is lost by  $N_f^I$. The lower boundary where the fixed point of the two loop beta function is lost is indicated by $N_f^{III}$ while the loss of large distance conformality estimated via the SD induced gap equation is indicated by $N_f^{II}$.  It is important to note that $N_f^{I},N_f^{II},$ and $N_f^{III}$ all denote a given number of Dirac fermions. For a given matter representation $R$ of $\mathrm{SU}(N_c)$ we have \cite{Dietrich:2006cm}:
 \begin{align}
 N_f^I[R]=\frac{11}{4}F_1[R], ~~
 N_f^{II}[R]=F_1[R]\frac{17 + 66F_2[R]}{10 + 30F_2[R]}, ~~
 N_f^{III}[R]=F_1[R]\frac{17}{10 + 6F_2[R]},
 \label{ladder}
 \end{align}
 with 
 \begin{align}
 F_1[R]=\frac{d[G]C_2[G]}{d[R]C_2[R]} \ , \qquad 
 F_2[R]=\frac{C_2[R]}{C_2[G]}
\end{align}
In these expressions $G$ denotes the adjoint representation and the $F$ functions hold information on the group-structure of the theory. For the two-index symmetric $S_2$ and two-index antisymmetric $A_2$ representations with a set of Dirac fermions one has
\begin{align}
 F_1[S_2/A_2]_\dia=\frac{2N_c}{N_c \pm 2} = 2 \mp \frac{4}{N_c} + \frac{8}{N_c^3} + {\cal O}(N_c^{-3})  \ .\end{align}
 and 
 \begin{align}
 F_2[S_2/A_2]&=1 \pm\frac{1}{N_c}-2\frac{1}{N_c^2}.
\label{Fs}
\end{align}
For the adjoint representation with Dirac fermions we have $F_1[G] = F_2[G] =1$. Note that the adjoint and two-index symmetric theories coincide in the small $N_c = 2 $ limit. Also in the small $N_c$ limit the two-index symmetric and antisymmetric theories diverge maximally in their dynamics. The divergence of the antisymmetric representation at $N_c=2$ is due to the fact that the representation becomes the singlet representation and decouples from the theory. 

At large $N_c$ the two-index symmetric and antisymmetric theories converge with the limiting values $F_1[S_2/A_2]= 2 F_2[S_2/A_2] =2$.  As expected in the large $N_c$ limit $F_1[S_2/A_2]$, which contains information about the dimension of the fermion representation, is twice the value for the adjoint representation $F_1[G]$.

It is instructive now to consider the leading $1/N_c$ corrections by inserting \eqref{Fs} into \eqref{ladder}. We find 
\begin{eqnarray}
 N_f^I[S_2/A_2] & = & 
 N_{wf}^I[G]\left( 1 \mp \frac{2}{N_c} + {\cal O}(N^{-2}_c)\right) \ , \\
    N_f^{II}[S_2/A_2] &= &N_{wf}^{II}[G] \left( 1 \mp \frac{649}{332 N_c} + {\cal O}(N^{-2}_c)\right) \ , \\
   N_f^{III}[S_2/A_2] &=  &N_{wf}^{III}[G] \left( 1 \mp \frac{19}{8N_c} + {\cal O}(N^{-2}_c)\right) \ , 
 \end{eqnarray} 
 with $N_{wf}[G]$ denoting the corresponding number of adjoint Weyl fermions
 \begin{eqnarray}
  N_{wf}^I[G] = \frac{11}{2} \ , \qquad  N_{wf}^{II}[G] = \frac{83}{20} \ , \qquad N_{wf}^{III}[G] = \frac{17}{8} \ .
 \end{eqnarray}
All three numbers $N_{wf}^{I}[G], N_{wf}^{II}[G]$ and $N_{wf}^{III}[G]$ are $N_c$ independent. As we decrease the number of colors we observe that the conformal window of the two-index symmetric (antisymmetric) theory  stays below (above) the corresponding conformal window of the adjoint theory. 
 
Since for $N_c=2$ the two-index symmetric representation is, by construction, equivalent to the adjoint representation it is also useful to formally expand around $N_c= 2$ 
\begin{eqnarray}
 N_f^I[S_2] &  = & 
 \frac{ N_{wf}^I[G]}{2}\left( 1 +\frac{( N_c  -2) }{4}  - \frac{(N_c - 2 )^2}{16}+  \cdots \right) \ , \\ 
   N_f^{II}[S_2] &= & \frac{N_{wf}^{II}[G]}{2} \left( 1 + \frac{347}{664}\frac{(N_c-2)}{2}  - \frac{1553}{5312}\frac{(N_c-2)^2}{4}+ \cdots \right) \ , \\
  N_f^{III}[S_2] &= & \frac{N_{wf}^{III}[G]}{2} \left( 1 + \frac{5}{16}\frac{(N_c-2)}{2} +\frac{17}{256}\frac{(N_c-2)^2}{4}+ \cdots \right) \ .
 \end{eqnarray}  
As we move to larger values of $N_c$ with respect to $N_c=2$ the boundary of the conformal window moves upwards. This aligns with the observation above that as we decrease the number of colors $N_c$ from the large $N_c$ limit the boundary of the conformal window moves to smaller values of $N_f[S_2]$. 
 
\begin{figure}[htbp]
\begin{center}
\includegraphics[width=0.8\textwidth]{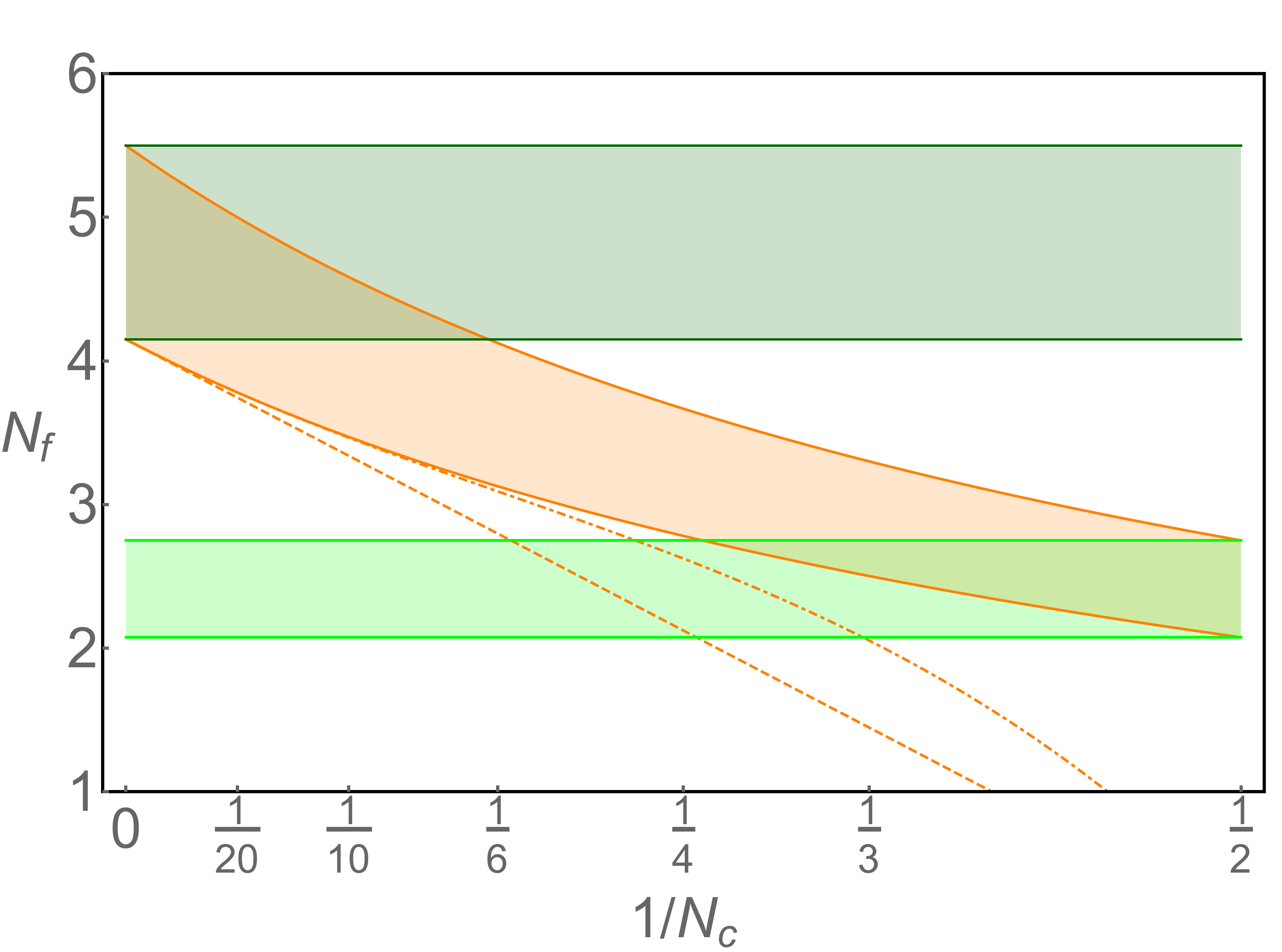}
\end{center}
\hspace{0.5cm}
\caption{The ladder conformal window for the two-index symmetric representation phase diagram (orange) and the adjoint representation (light green) in terms of the $N_f$ Dirac flavours. In dark green we show again the adjoint conformal window but where now $N_f$ is reinterpreted as the number of Weyl fermions $N_{wf}$.   For each window the upper bound(lower) corresponds to $N^I_f (N^{II}_f)$. The orange dashed curve corresponds to the $1/N_c$ expansion of $ N^{II}_f[S_2]$ up to and including ${\cal O}(N_c^{-1})$ while the orange dot-dashed curve corresponds to the $1/N_c$ expansion of $N^{II}_f[S_2]$ up to and including ${\cal O}(N_c^{-3})$. 
 }\label{CWL}
\end{figure}

We summarise in Fig.~\ref{CWL} the conformal window as estimated via the ladder approximation for the two-index symmetric and adjoint representations in terms of the number of Dirac flavours. The conformal window of the two-index symmetric theory is orange while the conformal window for the adjoint theory is light green. The dark green phase on the other hand is obtained by taking the adjoint conformal window and reinterpreting the number of adjoint Weyl fermions as Dirac flavors in the manner spelled out above. Hence via this reinterpretation at large $N_c$ it should be clear that the conformal window of the adjoint theory coincides with the conformal window of the two-index symmetric theory. 

For each window the upper (lower) bound corresponds to $N^I_f (N^{II}_f)$. The orange dashed curve corresponds to the $1/N_c$ expansion of $ N^{II}_f[S_2]$ up to and including ${\cal O}(N_c^{-1})$ while the orange dot-dashed curve corresponds to the $1/N_c$ expansion of $N^{II}_f[S_2]$ up to and including ${\cal O}(N_c^{-3})$. From the figure it is clear that the conformal window of the two-index symmetric theory is well described using the large $N_c$ limit only at very large $N_c$ since one needs at least ${\cal O}(N_c^{-3})$ corrections to arrive at $N_c \sim 6$ and new orders are needed to arrive at $N_c =3$. In fact, the expansion around $N_c=2$ converges more rapidly for smaller number of colors. 

Another way to estimate the conformal window is via perturbation theory to the maximal known loop order. Specifically one can employ the beta function of the gauge coupling and the anomalous dimension of the mass which in the modified minimal subtraction scheme, $\overline{\text{MS}}$, are known to four loop order \cite{vanRitbergen:1997va,Vermaseren:1997fq}. An investigation of the conformal phase to this loop order can be found in \cite{Pica:2010xq,Ryttov:2010iz}. Similar investigations in the modified regularisation invariant, RI', and minimal momentum subtraction, mMOM, schemes can be found in \cite{Ryttov:2014nda}. 

First one should evaluate the anomalous dimension at the zero of the beta function. Then by setting the lower boundary of the conformal window to be where the anomalous dimension reaches unity allows for an estimate of the critical number of flavors as a function of the number of colors. 

\begin{figure}[htbp]
\begin{center}
\includegraphics[width=0.6\textwidth]{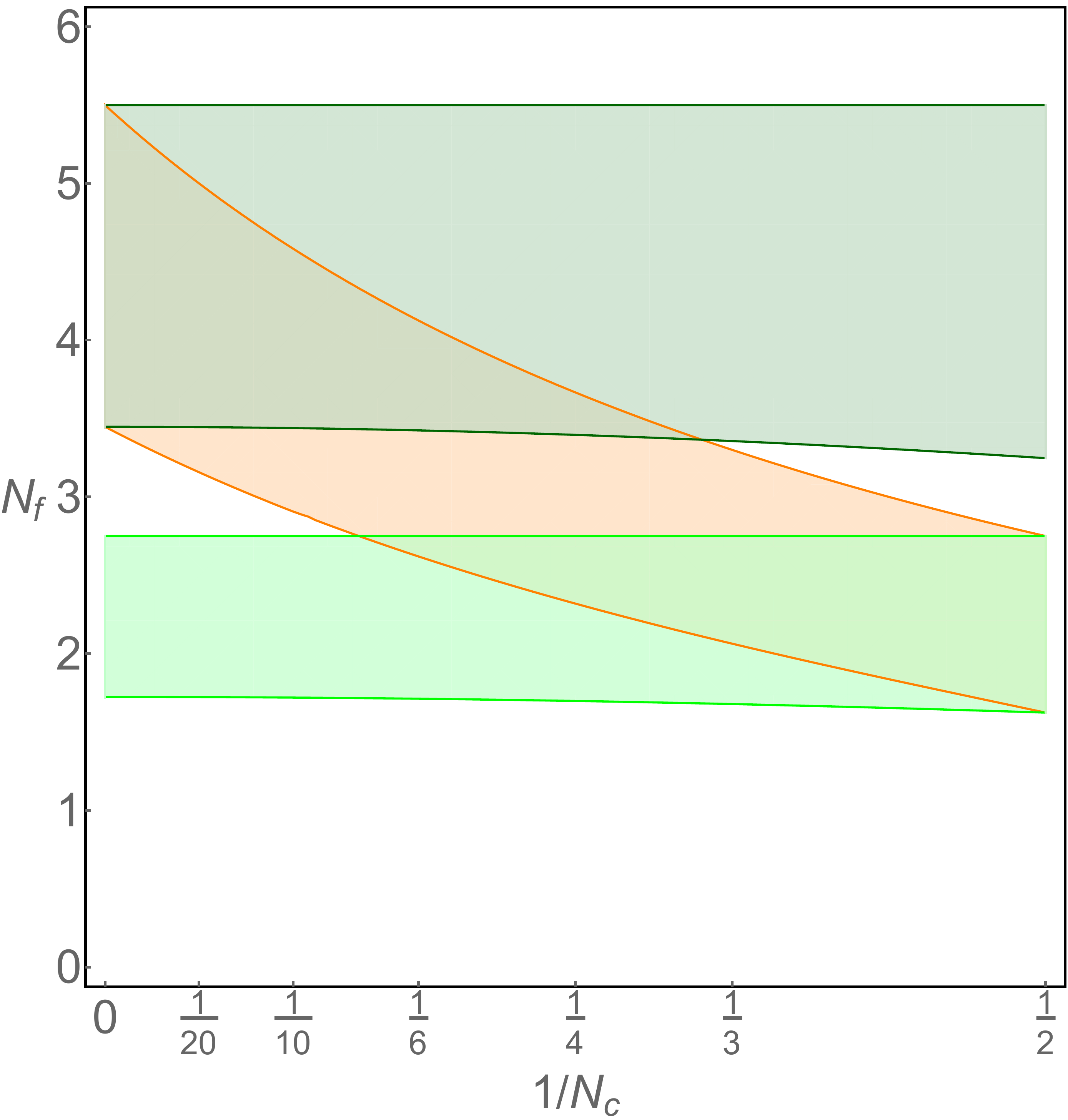}
\end{center}
\hspace{0.5cm}
\caption{The conformal window calculated using the four loop beta function and anomalous dimension of the mass. The boundary is set by the anomalous dimension reaching unity. The orange band is the conformal window for the two-index symmetric theory while the light green band is the conformal window for the adjoint theory both in terms of the number of Dirac flavors as a function of the inverse number of colors $1/N_c$. The dark green band is obtained by reinterpreting the number of adjoint Weyl fermions as the number of Dirac flavors.}\label{4loops}
\end{figure}

At two and three loops both the beta function and anomalous dimension of the mass only depend on the trace normalization factor and quadratic Casimirs for the various representations. This implies that the bound for the conformal window for the adjoint theory is independent of $N_c$. This is similar to the above estimate of the conformal window in the ladder approximation. However at four loops higher order group invariants enter the beta function and the anomalous dimension  \cite{vanRitbergen:1997va,Vermaseren:1997fq}. This has the effect of inducing an $N_c$ dependence in the associated estimate of the conformal window for the adjoint theory.  This dependence is very mild since it does not enter until the four loop level and we expect it to disappear again once the exact conformal window is known. These observations can be seen explicitly in Fig. \ref{4loops} where we present the numerical results. 

The orange band is the conformal window for the two-index symmetric theory while the light green band is the conformal window for the adjoint theory both in terms of a number of Dirac flavors. As can be seen for the adjoint theory there is a very mild dependence on $N_c$. The dark green band is the conformal window also for the adjoint representation but in this case $N_f$ counts the number of adjoint Weyl fermions. Again via this reinterpretation we see that the conformal window of the adjoint theory is equivalent to the conformal window of the two-index symmetric theory at large $N_c$.

The conformal windows estimated via the four-loop approximation are larger than the ones estimated via the previous SD. This happens because  the SD estimates involve only the two-loop beta function.

 \section{Concluding with non-perturbative results with(out) lattice }
 \label{section3}
We can now discuss further the nonperturbative results that extend beyond the ladder or the four-loop approximations by using large $N_c$ equivalence \cite{Corrigan:1979xf,Kiritsis:1989ge,Armoni:2004ub}. The equivalence resulted, among other things, in intriguing relations between $N=1$ supersymmetric Yang-Mills  and nonsupersymmetric two-index theories \cite{Armoni:2004ub}. The associated effective theory was constructed in \cite{Sannino:2003xe}.  The proof can be obtained  by means of the general expansion in terms of Wilson lines and is valid as long as no extra phase transitions separate the theories
\cite{Sannino:2005sk,Unsal:2006pj}. Hence the (exact) limiting $N_f^C$ for a conformal behaviour agrees for the adjoint, symmetric, and antisymmetric representation
at large $N_c$ provided that each Dirac flavors of the two-index theories is mapped in a single Weyl fermion of the adjoint theory. The general $1/N_c$ corrections are not known but we have been able to estimate them in the SD and four-loops approximation and shown that they are relevant. De facto the size of the corrections forbid simplistic extrapolations to small values of $N_c$. Although we considered explicitly the two-index symmetric representation the same holds true for the antisymmetric case.   

Nevertheless, based purely on the counting of degrees of freedom we can still arrive at the following useful inequalities at finite $N_c$
\begin{align}
 N_f^C[S_2] \leq N_{wf}^C[G]\leq N_f^C[A_2] \ ,
\label{eq:estimate}
\end{align}
for the limiting number of fermions where the conformal window starts. The inequalities hold because the effective number of adjoint Weyl  fermions is always smaller (larger) than the number of  two-index antisymmetric (symmetric) Dirac fermions. These constraints on the boundaries of the conformal windows hold true in our estimates above.  Therefore starting at large $N_c$ with $N_{wf}^C[G] = N_f^C[S_2]$ 
one expects the critical number of flavours  to move towards $N_{wf}^C[G]= \frac{1}{2}N_f^C[S_2]$ at $N_c=2$. The SD approximated value of  $N^C_f$ was indicated above by  $N^{II}_f$.

Another relevant point is that we expect the size of the adjoint conformal window not to depend on the number of colors. This is an exact statement for the upper bound, i.e. the value where we loose asymptotic freedom, and within the ladder approximation it is also true for the lower bound.  At four loops, as already noted, a small dependence appears which, however, can be neglected for all practical purposes and we expect it to disappear in the full nonperturbative treatment. 
%
%

These results allow for interesting crosschecks among different nonperturbative investigations of the conformal window. One well-established tool for these investigations are numerical lattice simulations that have already started to explore the dynamics of gauge theories with matter in higher dimensional representations.

We can, therefore, provide salient information on the conformal window and dynamics of the two-index theories by means of fermions in the adjoint representation at small $N_c$.  These results add value to lattice simulations of adjoint matter at small number of colors \cite{Catterall:2007yx,Catterall:2008qk,Catterall:2009sb,Hietanen:2009zz,DelDebbio:2010hx,DelDebbio:2010hu,Athenodorou:2015fda}.

Consider the case of two Dirac adjoint fermions of $\mathrm{SU}(2)$ which 
is believed to possess large scale conformality \cite{Catterall:2007yx,Catterall:2008qk,Catterall:2009sb, Hietanen:2009zz,DelDebbio:2010hx}. According to the arguments above this implies that {\it  at large $N_c$, the theories featuring  four Dirac fermions in the two-index representations display large distance conformality. Furthermore as we decrease the number of colors it remains conformal until $N_c = 16/3 $ where the theory looses asymptotic freedom. }  

Another  nonperturbative result is that {\it the conformal window for the two-index symmetric representation doubles in size from $N_c=2$ to infinity.}  This is so because when the number of colors drops to $N_c=2$  the two-index symmetric theory becomes the adjoint with matching Dirac fermions.  

We further note that recent studies hint to a potential conformal behaviour of the theory with one Dirac flavour in the adjoint representation  \cite{Athenodorou:2015fda}. It is still too premature to draw firm conclusions, however,  if further studies confirm the conformal behaviour it would demonstrate that the conformal window for the symmetric representation, for any $N_c$, starts below two Dirac flavours. This would therefore add independent insight on the (non) conformal faith of the phenomenologically relevant  $\mathrm{SU}(3)$ theory with two Dirac flavours   \cite{Fodor:2015zna,Hasenfratz:2015ssa}, because for any $N_c$  the theory would be conformal.

In conclusion by using large and small $N_c$ arguments as well as the $N_c$ independence of the conformal window for adjoint matter we deduced nonperturbative information on the conformal dynamics of theories with two-index    representations.  Consequently first principle lattice results of adjoint matter already provide valuable predictions for the conformal window and dynamics of two-index theories and further lead to important cross-checks and constraints among different theories.

 \section*{Acknowledgements}
CP$^3$-Origins is partially supported by the Danish National Research Foundation under the grant DNRF:90.

\end{document}